\documentclass[letterpaper, 10 pt, conference]{IEEEtran}
\usepackage{hyperref}
\usepackage{amsmath}
\usepackage{amssymb}

\usepackage{algorithmicx}
\usepackage{algorithm}
\usepackage{algpseudocode}

\usepackage{graphicx}
\usepackage{tikz}

\begin{document}
%
\title{Reversible Kalman Filter for state estimation with Manifold}

\author{
\IEEEauthorblockN{Svyatoslav Covanov}
\IEEEauthorblockA{
IRL2958 GT-CNRS\\
GeorgiaTech Europe - Metz, France\\
svyatoslav.covanov@polytechnique.org}\\
\and
\IEEEauthorblockN{C\'edric Pradalier}
\IEEEauthorblockA{
IRL2958 GT-CNRS\\
GeorgiaTech Europe - Metz, France\\
cedric.pradalier@georgiatech-metz.fr}
}
\markboth{Journal of \LaTeX\ Class Files,~Vol.~14, No.~8, August~2015}%
{Shell \MakeLowercase{\textit{et al.}}: Bare Demo of IEEEtran.cls for IEEE Journals}
%



\maketitle
\newcommand\submittedtext{%
  \footnotesize This work has been submitted to the IEEE for possible publication. Copyright may be transferred without notice, after which this version may no longer be accessible.}

\newcommand\submittednotice{%
\begin{tikzpicture}[remember picture,overlay]
\node[anchor=south,yshift=10pt] at (current page.south) {\fbox{\parbox{\dimexpr0.65\textwidth-\fboxsep-\fboxrule\relax}{\submittedtext}}};
\end{tikzpicture}%
}

\begin{abstract}
This work introduces an algorithm for state estimation on manifolds within the framework of the Kalman filter. Its primary objective is to provide a methodology enabling the evaluation of the precision of existing Kalman filter variants with arbitrary accuracy on synthetic data, something that, to the best of our knowledge, has not been addressed in prior work.

To this end, we develop a new filter that exhibits favorable numerical properties, thereby correcting the divergences observed in previous Kalman filter variants. In this formulation, the achievable precision is no longer constrained by the small-velocity assumption and is determined solely by sensor noise.

In addition, this  new filter assumes high precision on the sensors, which, in real scenarios require a detection step that we define heuristically, allowing one to extend this approach to scenarios, using either a 9-axis IMU or a combination of odometry, accelerometer, and pressure sensors. The latter configuration is designed for the reconstruction of trajectories in underwater environments.
\end{abstract}

\submittednotice

%
\IEEEpeerreviewmaketitle

\section{Introduction}

%
%
%
%

{T}{he} present work is motivated by applications in the routine inspection of large metallic structures, such as ship hulls, in underwater environments. The general scenario involves deploying differential-drive robots equipped with acoustic sensing techniques to inspect ship surfaces. The specific problem addressed in this paper is the localization of the robot. To this end, we investigated different sensor configurations, beginning with a theoretical analysis. A general discussion of state estimation in robotics can be found in~\cite{10.5555/3165227}.

A related effort is presented in~\cite{starbuck:hal-03445976}, where a method derived from the Invariant Extended Kalman Filter (IEKF) as described in~\cite{annurev:/content/journals/10.1146/annurev-control-060117-105010} was applied to localize a magnetic crawler on a ship hull. That work emphasized the development of a rigorous mathematical framework, relying on wheel encoders for the prediction step and ultra-wideband measurements for the correction step.
This approach followed the idea to implement consitent variant of Kalman filters with the property of a manifold, which has been described in~\cite{phdthesisReady,Barrau2015AnEA}

However, this approach do not fit to sensors typically embedded in an IMU—such as gyroscopes, accelerometers, or magnetometers—is challenging. These sensors require a three-dimensional state representation, whereas wheel encoders encode a two-dimensional state vector (which, although embedded in 3D, lies on a 2D differentiable manifold). The central difficulty lies in defining a consistent projection of IMU measurements onto the 2D manifold while preserving the covariance structure of the Kalman filter. For instance, if the robot orientation provided by the gyroscope is constrained to remain approximately normal to the manifold, then accelerometer and magnetometer measurements must be appropriately rotated so that the Kalman update remains consistent with the 2D state formulation.

This problem gives rise to two specific challenges. First, the accelerometer measures both gravity and external accelerations, making their separation nontrivial. Several recent approaches address this by applying signal processing techniques to filter certain components of the accelerometer signal~\cite{Lee2012EstimationOA,Widodo2016}. Second, magnetometer readings are often affected by magnetic disturbances, which are common near ship hulls. One mitigation strategy is to incorporate pressure sensors as complementary information.

Beyond filtering strategies, alternative approaches exploit system symmetries, giving rise to modern equivariant frameworks~\cite{Mahony2020,Mahony2022,FORNASIER2025112495}. This line of work builds on decades of research into symmetry-preserving estimation, notably through the development of the invariant Kalman filter~\cite{Bonnabel2009}.

Our approach is complementary to these efforts and focuses on the \textit{testability} of existing Kalman filter variants, where numerical precision on synthetic data depends on data variations. Rather than embedding Kalman filters within a general symmetry-based framework, we focus on the symbolic treatment of constraints arising from overdetermined sensor configurations (i.e., when more measurements are available than typically required). Our emphasis is on numerical precision, studied primarily on synthetic data. This has led to the design of a filter that, by exploiting prior knowledge of the surface, can compensate for sensor deficiencies. A key theoretical result is that, in the absence of magnetic disturbances and under surface-constrained motion, external accelerations can be eliminated with arbitrary precision. Combined with gyroscopic measurements, this enables broader applications such as trajectory estimation for vehicles on flat surfaces (e.g., autonomous ground vehicles) or nearly flat surfaces (e.g., human motion analysis). Another application is underwater trajectory reconstruction, where the robot orientation is assumed to remain normal to the surface, allowing pressure sensors to replace magnetometers.

To formalize this approach, we compare the Multiplicative Kalman Filter (MEKF) with our method and introduce the mathematical property satisfied by our filter, which we refer to as reversibility. This property characterizes desirable behavior in the considered context. We conclude the theoretical discussion by presenting the algorithmic construction of the proposed filter, which satisfies this property. Then, we describe at which conditions this filter can be transfered to real data. In particular, one needs to define a detection step, to capture the moment when the input sent by the sensors can be corrected with our approach. This detection step is a heuristic that enable its practical use in experimental settings.

The paper is structured as follows, using $9$-axis IMU sensors as the main example. 
Section~\ref{sec:backtheory} introduces the notation and provides a theoretical overview of quaternion properties. 
Section~\ref{sec:mekf} presents the Multiplicative Extended Kalman Filter (MEKF), used as a reference throughout this work, with an emphasis on its numerical properties. 
Section~\ref{sec:revmekf} details the theoretical approach we propose, while Section~\ref{sec:evalmekf} evaluates its performance on both synthetic and real data. 
Section~\ref{sec:discussions} discusses the main applications and limitations of our method. 
Finally, the Appendix outlines how our approach can be extended to sensor configurations combining odometry, accelerometer and pressure sensors instead of IMUs.
The code for this project is available at the following repository: 
\href{https://github.com/knov836/revmekf}{https://github.com/knov836/revmekf}.

\section{Background Theory}
\label{sec:backtheory}

In this section, we introduce the necessary notations and mathematical background required for the formulation of our problem. Since our objective is to achieve arbitrary numerical precision in trajectory computation, a rigorous treatment of rotations and sensor models is essential.  

\subsection{Quaternions, Rotations, and Axis-Angle Representation}

We represent elements of $SO(3)$ (3D rotations) using quaternions. A detailed exposition of quaternion kinematics can be found in~\cite{DBLP:journals/corr/abs-1711-02508}.  

For a rotation $R \in SO(3)$ and a vector $v \in \mathbb{R}^3$, we denote by $R \cdot v$ the action of $R$ on $v$. Let $(0|v)$ denote the quaternion associated with $v$ by appending a fourth coordinate equal to zero. There exists a unit quaternion $q$ such that  
\[
q \,(0|v)\, q^{-1} = (0|R \cdot v),
\]  
establishing the correspondence between unit quaternions and rotations. We denote by $R(q)$ the rotation matrix associated with a unit quaternion $q$.  

Furthermore, quaternions admit a logarithmic map, providing a correspondence between unit quaternions and elements of $\mathbb{R}^3$:  
\[
\log(q) = u = \alpha \mathbf{u},
\]  
where $\mathbf{u}$ is the unit vector along the axis of rotation, and $\alpha$ is the rotation angle.

Conversely, the exponential map from $\mathbb{R}^3$ to $SO(3)$ is given by  
\[
\exp(u) = I_3 + \sin(\alpha)\,[\mathbf{u}]_\times + (1-\cos(\alpha))\,[\mathbf{u}]_\times^2,
\]  
where $[\mathbf{u}]_\times$ is the skew-symmetric matrix associated with $\mathbf{u}$.
The relation between logarithmic map of quaternion and Lie groups is described in~\cite{Sol2018AML}.

\subsection{IMU Measurement Model}

We consider a standard IMU composed of a gyroscope, an accelerometer, and a magnetometer. The gyroscope provides angular velocity measurements. For a quaternion $q_k$ representing orientation at time step $k$, the gyroscope yields a 3D vector $\boldsymbol{\omega}_k$ such that  
$$q_{k}^{\mathcal{G}} = q_{k-1}^{\mathcal{G}}\cdot \exp(\mathbf{\omega_k}\cdot dt)$$ and $$q_{k}^{\mathcal{R}} = \exp(-\mathbf{\omega_k}\cdot dt)\cdot q_{k-1}^{\mathcal{R}}$$
where $\mathcal{G}$ denotes the global frame and $\mathcal{R}$ the relative frame.  

The exponential map above provides the link between quaternions and the axis-angle representation of rotations.  

A global quaternion enables the transformation of sensor measurements from the relative frame to the global frame. For the accelerometer, this yields  
\[
q_{k}^{\mathcal{G}} \cdot \mathbf{A}_k^{\mathcal{R}} \cdot \big(q_{k}^{\mathcal{G}}\big)^{-1} = \mathbf{A}_k^{\mathcal{G}}.
\]  
In the absence of external accelerations, $\mathbf{A}_k^{\mathcal{G}}$ is a noisy estimate of the gravity vector.  

Similarly, for the magnetometer,  
\[
q_{k}^{\mathcal{G}} \cdot \mathbf{M}_k^{\mathcal{R}} \cdot \big(q_{k}^{\mathcal{G}}\big)^{-1} = \mathbf{M}_k^{\mathcal{G}},
\]  
where $\mathbf{M}_k^{\mathcal{G}}$ is a measurement of the Earth’s magnetic field.  

In practice, the available IMU outputs are $\boldsymbol{\omega}_k$, $\mathbf{A}_k^{\mathcal{R}}$, and $\mathbf{M}_k^{\mathcal{R}}$.  

\subsection{Integration of Gyroscopic Data}

The algorithms described in this work rely on quaternions and axis-angle representations to estimate the global orientation, which requires the integration of gyroscopic data.  

For $k > 0$, the relation between successive quaternions $q_{k+1}^{\mathcal{G}}$ and $q_{k}^{\mathcal{G}}$ is given by  
\[
\log\left(\left(q_{k}^{\mathcal{G}}\right)^{-1} \cdot q_{k+1}^{\mathcal{G}}\right) = \boldsymbol{\omega}_k \, \Delta t.
\]  

Equivalently,  
\[
q_{k+1}^{\mathcal{G}} = q_{0}^{\mathcal{G}} \cdot \prod_{i=0}^{k} \left( \left(q_{i}^{\mathcal{G}}\right)^{-1} \cdot q_{i+1}^{\mathcal{G}} \right).
\]  

This formulation corresponds to the discretization of the continuous-time differential equation~\cite{DBLP:journals/corr/Zhao16c}:  
\[
\dot{q}^{\mathcal{G}} = q^{\mathcal{G}} \cdot [\boldsymbol{\omega}(t)]_\times,
\]  
which leads, after discretization, to  
\[
q_{k+1}^{\mathcal{G}} = q_{k}^{\mathcal{G}} \cdot \exp\big(\boldsymbol{\omega}_k (t_{k+1}-t_k)\big).
\]  

\section{Multiplicative Extended Kalman Filter (MEKF)}
\label{sec:mekf}

In this section, we present a variant of the Kalman filter in which the mathematical objects involved in the computation are better adapted to the structure of the inner state and the constraint on the norm of quaternions. 
This variant has been described in~\cite{Maley2013MultiplicativeQE}.

\subsection{Structure of the algorithm}
In the case of IMU sensors, the Multiplicative Extended Kalman Filter (MEKF) takes as input a sequence of sensor measurements 
$\boldsymbol{\omega}_k$, $\boldsymbol{A}_k^{\mathcal{R}}$, and $\boldsymbol{M}_k^{\mathcal{R}}$.

At each sample $k$, MEKF updates an inner state vector $\hat{x}$ in two steps: a prediction step and an update step. The structure of $\hat{x}$ is a 6-dimensional vector, where the first three components correspond to the logarithm of a unit quaternion (axis–angle representation), and the last three components correspond to the gyroscope bias.

\begin{itemize}
\item \textbf{Prediction step:} compute $q$ using gyroscope data, and update
\begin{align*}
    \hat{x}[1:3] &\leftarrow \log(q), \\
    \hat{x}[4:6] &\leftarrow \hat{x}[4:6].
\end{align*}

\item \textbf{Update step:} in this step, MEKF corrects the inner state vector using accelerometer and magnetometer information. In static conditions, one can recover a unique quaternion $q$ from $\boldsymbol{A}_k^{\mathcal{R}}$ and $\boldsymbol{M}_k^{\mathcal{R}}$, using the identity

%
%
$ q_{k}^{\mathcal{G}}\cdot \left(\mathbf{A}_k^{\mathcal{R}}\otimes \mathbf{M}_k^{\mathcal{R}}\right)\cdot \left(q_{k}^{\mathcal{G}}\right)^{-1} = \left(q_{k}^{\mathcal{G}}\cdot \mathbf{A}_k^{\mathcal{R}}\left(q_{k}^{\mathcal{G}}\right)^{-1}\right) \otimes \left(q_{k}^{\mathcal{G}}\cdot \mathbf{M}_k^{\mathcal{R}}\left(q_{k}^{\mathcal{G}}\right)^{-1}\right),$
where $\otimes$ denotes the cross product.
In other words, the knowledge of the rotation of two orthogonal vectors is sufficient to determine a third one, and thus uniquely defines a quaternion.
\end{itemize}

\begin{algorithm}
\caption{Kalman Filter: Prediction and Update}\label{alg:Kalman}
\begin{algorithmic}[1]

	\Require Previous state estimate $\hat{x}_{k-1}$, Quaternion $q$, Bias $B$, Covariance $P_{k-1}$, Gyroscope $\boldsymbol{\omega} = (p, q, r)$, Accelerometer $\boldsymbol{A}$, Magnetometer $\boldsymbol{M}$, Time step $dt$
\Ensure Updated state estimate $\hat{x}_k$, Covariance $P_k$

\Statex $\hat{x}_{k|k-1} \leftarrow$\textbf{Prediction}($\hat{x}_{k-1},\boldsymbol{\omega}$)
\Statex $\hat{x}_{k} \leftarrow$ \textbf{Update}($\hat{x}_{k|k-1},\boldsymbol{A},\boldsymbol{M}$)

\State \Return $\hat{x}_k, P_k$

\end{algorithmic}
\end{algorithm}

\subsection{Prediction step}
The prediction relates to an integration operation on the gyroscopic input. The procedure is well described in~\cite{Maley2013MultiplicativeQE}.

\begin{algorithm}
\caption{EKF Prediction Step for Quaternion Orientation}\label{alg:Prediction}
\begin{algorithmic}[1]

	\Require Quaternion $q_{k-1}$, Bias $B_{k-1}$, Covariance $P_{k-1}$, Process noise $Q_k$, Gyroscope $\boldsymbol{\omega} = (p, q, r)$, Time step $dt$
\Ensure Updated $q_k$, $B_k$, $P_k$
	\State $[\omega_x, \omega_y, \omega_z] \gets [p - B_{k,0}, q - B_{k,1}, r - B_{k,2}]$
    \State $\Delta q \gets \exp([\omega_x, \omega_y, \omega_z] \cdot dt)$
	\State $q_{k|k-1} \gets q_{k-1} \cdot \Delta q$ \Comment{Right multiplication because global frame}
    \State Initialize $F \gets 0_{6 \times 6}$, $\Phi \gets I_6$
    \State $F_{1:3,1:3} \gets -\text{skew}([\omega_x, \omega_y, \omega_z])$
    \State $F_{1:3,4:6} \gets -I_3$
    \State $\Phi \gets I_6 + F \cdot dt$ \Comment{If $dt$ is large, higher-order terms may be needed}
	\State $P_k \gets \Phi \cdot P_{k-1} \cdot \Phi^\top + Q_k$
    \State \Return $q_{k|k-1}, B_{k-1}, P_k, \Phi$
\end{algorithmic}
\end{algorithm}

Given gyroscopic data $[\omega_{x},\omega_{y},\omega_{z}]$, the integration computes
\[
	q_{k-1} \cdot \exp([\omega_{x},\omega_{y},\omega_{z}] \cdot  dt),
\]
which corresponds to the quaternion $q_{k|k-1}$.

The covariance is computed by using a linearization of the exponential map, corresponding to the matrix $\Phi$.

\subsection{Update step}

The update step is described by Algorithm~\ref{alg:Update}. Its construction can be seen as a compromise between the inner state computed by the prediction step and the state deduced from accelerometer and magnetometer measurements.

In the multiplicative Kalman variant, the inner vector state is interpreted as a perturbation $\eta$ on the quaternion representing the orientation. Thus, the update step applies
\[
q_k \gets q_{k|k-1} \cdot \exp(\eta).
\]

\begin{algorithm}
\caption{EKF Update Step}\label{alg:Update}
\begin{algorithmic}[1]

	\Require Predicted state $\hat{x}_{k|k-1}$, Quaternion $q_{k|k-1}$, Bias $B_{k|k-1}$, Covariance $P_{k|k-1}$, Measurement noise $U_k$, Accelerometer $\boldsymbol{A}$, Magnetometer $\boldsymbol{M}$
\Ensure Updated state $\hat{x}_k$, Covariance $P_k$

    \State $z_k \gets (\boldsymbol{A}, \ \boldsymbol{M})$
    \State $y_k \gets z_k - H_k \hat{x}_{k|k-1}$ \Comment{Innovation}
    \State $S_k \gets H_k P_{k|k-1} H_k^\top + U_k$ \Comment{Innovation covariance}
    \State $K_k \gets P_{k|k-1} H_k^\top S_k^{-1}$ \Comment{Kalman gain}
    \State $\hat{\eta}_k \gets K_k y_k$ \Comment{Quaternion perturbation}
    \State $q_k \gets q_{k|k-1} \cdot \exp(\hat{\eta}_k)$
    \State $\hat{x}_k \gets \log(q_k)$
    \State $P_k \gets (I - K_k H_k) P_{k|k-1}$

    \State \Return $\hat{x}_k, P_k$

\end{algorithmic}
\end{algorithm}

In this algorithm, one has to define the measurement noise covariance $U_k$ and the transition matrix $H_k$.  

The noise covariance can be chosen diagonal, defined by 
\[
U_k = \mathrm{diag}(u_0,u_1,u_2,u_3,u_4,u_5),
\] 
where the $u_i$ are tuning parameters of the filter.  

The transition matrix $H_k$ can be divided in $3\times 3$ blocks:
\[
H_k = \begin{bmatrix}
    [R(q)\boldsymbol{g}]_{\times} & 0 \\
	[R(q)\boldsymbol{b}]_{\times} & 0
\end{bmatrix},
\]
where $[v]_{\times}$ denotes the skew-symmetric matrix associated to vector $v$, $R(q)$ is the rotation matrix associated with quaternion $q$, $\boldsymbol{g}$ is the gravity vector, and $\boldsymbol{b}$ is the magnetic field vector in the earth frame.

\subsection{Multiplicative residual}

In the previous variant, one defines a residual $y_k$ as
$$y_k \gets z_k  -H_k \hat{x}_{k|k-1}.$$

One can compute the following approximation of the rotation $R(q)$ associated to $q$
$$R(\delta q) = I-2[q_{2:4}]_\times$$
which leads to
$$R(q) \approx R(\hat{q}) \cdot (I+[\eta]_\times).$$

The consequence of this relation is that one can relate an input vector $\mathbf{A}$ with the rotation of a vector $\Vec{g} = [0, 0, g]$ in the following way:
$$\mathbf{A} -\hat{\mathbf{A}} = -[\eta]_\times R(q) \Vec{g}.$$
This formula gives the transition matrix $H$ under the following form
$$H = \begin{pmatrix}
    [R(q) \Vec{g}]_\times & 0\\
    [R(q) \Vec{b}]_\times & 0\\
\end{pmatrix}.$$

From this method, one can generalize the computation of transition matrix from different type of residual. In particular, it is possible to select a multiplicative residual, defined as
$$\mathbf{A}\otimes \hat{A}$$
for the acceleration. This approach has been described in~\cite{zanetti:mult_residual}.
Using the same formula for the magnetometer,
one gets the following transition matrix $H$:
$$H = \begin{pmatrix}
    I-(R(q) \Vec{g}) \cdot (R(q) \Vec{g})^T & 0\\
    I-(R(q) \Vec{b}) \cdot (R(q) \Vec{b})^T & 0\\
\end{pmatrix}.$$

\subsection{Behaviour on synthetic data}

In this section, we analyze the behaviour of the previously described algorithm on synthetic data, corresponding to a random $3D$ trajectory. This setting allows us to compare the performance of the algorithms in terms of the loss induced by numerical imprecision.

\begin{figure}[h!]
  \centering
  \includegraphics[width=0.33\textwidth]{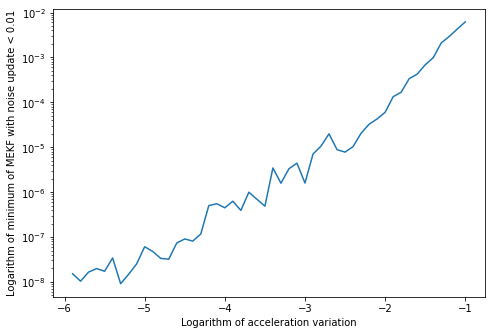} 
  \caption{Evolution of $\max(|p_{\text{MEKF}} - p_{\text{truth}}|)$ as a function of logarithm of average acceleration variations during $dt=1/100s$.}
  \label{fig:evolution_pos}
\end{figure}

We use noise matrices $Q$ and $U$ with coefficients set to $10^{-2}$.  
As illustrated in Figure~\ref{fig:evolution_pos}, the smaller the variation of the acceleration, the smaller the final position error. Thus, the MEKF exhibits satisfactory behaviour in the case of low accelerations. However, its precision depends on the dynamics of the motion rather than solely on the quality of the data. 

From this, we conclude that the position estimated by the MEKF is intrinsically sensitive to variations in the accelerometer: the smaller these variations, the more accurate the estimate. In contrast, integrating the gyroscope alone yields a position estimate with a numerical error on the order of $N \cdot 10^{-40}$, where $N$ denotes the number of samples. By comparison, the error in the MEKF remains strongly dependent on the characteristics of the motion.

\medskip
We next analyze the behaviour of MEKF under different parameter settings by introducing a bias in the gyroscopic data. For acceleration variations ranging from $10^{-5}$ to $10^{-10}\,\mathrm{m.s^{-2}}$ during $dt$, and for a sampling frequency of $100\,\mathrm{Hz}$, we generated data with a bias up to $10^{-2}$.  

\begin{figure}[h!]
  \centering
  \includegraphics[width=0.33\textwidth]{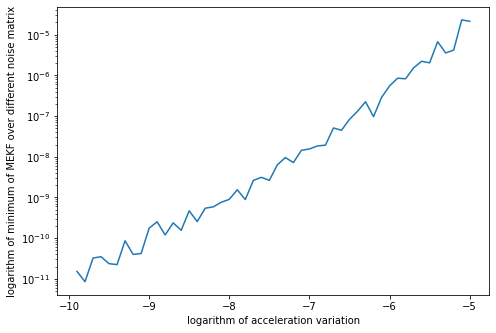} 
	\caption{Evolution of $\log(\max(|p_{\text{MEKF}} - p_{\text{truth}}|))$ for different noise matrices down to $10^{-10}$, as a function of logarithm of acceleration variations.}
  \label{fig:evolution_pos_bias_mekf}
\end{figure}

Figure~\ref{fig:evolution_pos_bias_mekf} reports the logarithm of the minimum precision achieved by MEKF depending on the logarithm of the average acceleration variation during $dt$ when the norm of the coefficients of the update noise matrix vary between $10^{-1}$ and $10^{-10}$. A clear linear dependency with respect to the acceleration variation can be observed.

Depending on the acceleration variation, the minimum precision achieved by MEKF changes accordingly. One can estimate that this precision is always greater than the average acceleration variation occurring during the sampling interval $dt$.

\section{Reversible filter}
\label{sec:revmekf}
\subsection{Theoretical aspect}
We define in this section the mathematical concept that will be useful to compare various filters. The main idea is that for some action on measurements, one can define a reversible filter, which can be thought as a filter allowing one to go backward.

Let $G$ be a group, which can be thought as $SO(3)$.
Let $\mathcal{M} = G\times V$ be the the set of measurements, where $V$ is a vector space, on which one can define an action of $G$.
We define the following left action of $G$ on $\mathcal{M}$:
$$h * (v,w) = (h\cdot v,h\cdot w).$$

A filter can be defined as a function $f$ mapping $(\mathcal{S},\mathcal{M})$ to $\mathcal{S}$, where $\mathcal{S}$ is the set of state vectors.

One can define a \textit{$G$-reversibility} property:
$$\forall u, \forall m = (h,w), f(f(u,m),h^{-1}*(Id_G,w)) = u.$$

The \textbf{strong} \textit{$G$-reversibility} property for a filter described by the function $f$ can be defined as:
$$\forall \mathbf{\epsilon},\forall u, \forall m, f(f(u,m),h^{-1}*(Id_G,w)+\mathbf{\epsilon}) = u+O_f(\mathbf{\epsilon}).$$

In other terms, a filter is strongly \textit{$G$-reversible} if an error of amplitude $\epsilon$ on the measurements leads to a bounded error of amplitude $O_f(\epsilon)$ depending only on the function $f$. 

We define a set of filters for which we compare a new approach. In particular, we demonstrate that the Multiplicative Kalman Filter, is not reversible.

\subsection{MEKF is equivariant but not reversible}

One can easily demonstrate the non reversibility of MEKF by using the following counter-example.
Assume that one is using IMU sensors, with no noise on gyroscope, accelerometer and magnetometer.
At time $t=0$, assume that there is no rotation, thus, we consider measurements of type
$m=(Id_{SO(3)},(\mathbf{A},\mathbf{M}))$.

One can consider the case where $\mathbf{M} = [0,1,0]$ and $\mathbf{A} = [0,0,1] + y\cdot [0,1,0]$. Thus, in our example, the movement is directed in $y$ axis, and $y\cdot [0,1,0]$ is the external acceleration. For small values of $y$, the correction applied by MEKF tends to behave as
$$f([0,0,0],m) \approx [0,0,0],$$
which means that the axis-angle representation of the orientation is close to $0$.

For big values of $y$, the behaviour is
$$f([0,0,0],m) \otimes [1,0,0] \approx 0,$$
which means that the orientation deduced from MEKF is orthogonal to $[1,0,0]$.

In that case, considering reversibility property would mean studying the behaviour of
$$f(f([0,0,0],m),(Id_{SO(3)}, (\mathbf{A},\mathbf{M}))$$
which consists in applying again the previous transform on the state vector vector $f([0,0,0],m)$. The logarithm of the computed quaternion is colinear to $[1,0,0]$.

The reversibility property is orthogonal to the equivariant property described in~\cite{FORNASIER2025112495}. This last property corresponds to the existence of an equivariant filter design equivalent to MEKF. The effort put in that direction tries to capture the existent symmetries of MEKF. In our context, we propose a variation of MEKF that tries to compensante an asymmetry by using a relation between the surface knowledge and the accelerometer.

\subsection{Linear algebra in Kalman filter}

In this section, we formalize an intermediate stage inserted between the prediction and update steps. Its purpose is to derive the gravity vector from the available input data and, in turn, to substitute this vector for the raw accelerometer measurements in the update step of the MEKF.

Given the IMU measurements $(\mathbf{\omega},\mathbf{A},\mathbf{M},\Vec{n})$, one can decompose these measurements as
$m=(\mathbf{\omega},\mathbf{A}_g^R+\mathbf{A}_{\text{ext}}^R,\mathbf{M},\Vec{n})$, where $\mathbf{A}_g$ is the rotation of the gravity vector.

The idea that we use is that in that context, there exists a rotation $R$ such that
$$m = (\mathbf{\omega},R\cdot \Vec{g} + R\cdot \mathbf{A}_{\text{ext}}^G,R\cdot \Vec{b},\Vec{n}).$$
There is a finite number of rotation $R$ such that, given a position $\mathbf{p}$ and a speed $\mathbf{s}$ satisfying $\mathbf{p} \cdot \Vec{n} = 0$ (the position is on the surface), such that
$$(\mathbf{p}+\mathbf{s}\cdot dt -\Vec{g}\cdot (dt)^2 +R^{-1}\cdot \mathbf{A})\cdot \Vec{n} = 0$$
and
$$R^{-1} \mathbf{M} = \Vec{b}.$$
The reason of the finite number of rotations comes from the fact that $\Vec{b}\neq \Vec{n}$ (the magnetic field not colinear to the normal, which can be assumed without loss of generality) and from the fact that the logarithm of the rotation such that $R^{-1} \mathbf{M} = \Vec{b}$ are constrained to an ellipse in the plan equidistant to $\mathbf{M}$ and $\Vec{b}$. If $\mathbf{M} = \Vec{b}$, the possible logarithms are given by the rotation
of axis $\Vec{b}$, which form a line.

Since any rotation $R$ can be decomposed as the product of $R_{\Vec{b}}$ (rotation leaving $\Vec{b}$ invariant) and $R(\Vec{b} \to \mathbf{M})$ (rotation sending $\Vec{b}$ to $\mathbf{M}$), we do not lose generality by considering the previous singular case. Assuming that $\Vec{b} \neq \Vec{n}$, one can use the first condition to compute $2$ intersections with the plan orthogonal to $\Vec{n}$.
In conclusion, one can use the two conditions to compute a unique rotation, closest to the prediction step in MEKF. From this rotation, one deduce the measurements
$(\mathbf{\omega},\mathbf{A}_g^R,\mathbf{M},\Vec{n})$.

\subsection{Pseudo-code}

Adding a new step in Algorithm~\ref{alg:RevMEKF}, one obtains a reversible variant of Kalman filter. The reason is that each term of the measurements $m$
is only dependent on the orientation:
given $R$ such that $R\cdot \Vec{g} = \mathbf{A}$ and $R\cdot \Vec{b} = \mathbf{M}$, one has
$$h^{-1}*(Id,(\mathbf{A},\mathbf{M})) = (h^{-1},(h^{-1}R\cdot \Vec{g},h^{-1}R\cdot \Vec{b}))$$
and
$$f(f(u,(h,(\mathbf{A},\mathbf{M}))),h^{-1}*(Id_{SO(3)},(\mathbf{A},\mathbf{M}))) = u.$$

The consequence of the previous relation and the continuity of the linear algebra part is also the strong reversibility of this algorithm.
\begin{algorithm}
\caption{Reversible Kalman Filter: Prediction and Update}\label{alg:RevMEKF}
\begin{algorithmic}[1]

	\Require Previous state estimate $\hat{x}_{k-1}$, Quaternion $q$, Bias $B$, Covariance $P_{k-1}$, Gyroscope $\mathbf{\omega} = (p, q, r)$, Accelerometer $\boldsymbol{A}$, Magnetometer $\boldsymbol{M}$, Time step $dt$
\Ensure Updated state estimate $\hat{x}_k$, Covariance $P_k$

\Statex $\hat{x}_{k|k-1} \leftarrow$\textbf{Prediction}($\hat{x}_{k-1},\mathbf{\omega}$)
\Statex $\boldsymbol{A}_g \leftarrow$\textbf{LinAlg}($\hat{x}_{k|k-1},\boldsymbol{A}$,$\boldsymbol{M}$)
\Statex $\hat{x}_{k} \leftarrow$ \textbf{Update}($\hat{x}_{k|k-1},\boldsymbol{A}_g$,$\boldsymbol{M}$)

\State \Return $\hat{x}_k, P_k$

\end{algorithmic}
\end{algorithm}

\section{Evaluation of Rev-MEKF}
\label{sec:evalmekf}

We applied the proposed Rev-MEKF algorithm to synthetic data in order to demonstrate its improvements over MEKF, and in particular to show that its numerical precision is independent of the type of motion. This makes Rev-MEKF a suitable tool for testing the implementation of the update step in MEKF.

In addition, we evaluated a variation of Rev-MEKF on real data collected from a driving vehicle equipped with an IMU. Real data bring additional challenges, which we discuss in this section.

\subsection{Synthetic data}

We considered three experimental conditions: 
\begin{enumerate}
    \item no noise, 
    \item gyroscope bias of order $10^{-8}$ $\text{rad.s}^{-2}$ with no noise on the accelerometer, 
    \item accelerometer noise of order $10^{-15}m.s^ {-2}$ with noise on the gyroscope of order $10^{-8}$ $\text{rad.s}^{-2}$.
\end{enumerate}

In all our examples, the average variation of acceleration is less than $10^{-1}m.s^{-2}$.

Our first observation is that Rev-MEKF exhibits a significant improvement compared to MEKF. In the noise-free case, the loss of precision depends only on time and on the fixed numerical precision of the input sensors. The error remains bounded by a constant factor times $N\cdot 10^{-40}$, where $N$ is the number of samples.

\begin{figure}[h!]
  \centering
  \includegraphics[width=0.33\textwidth]{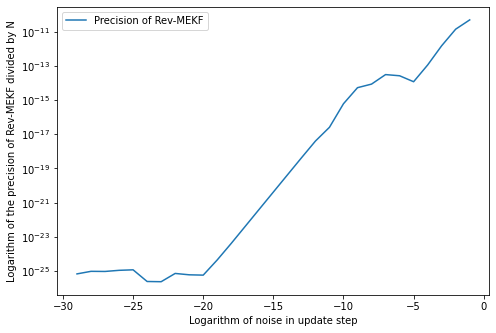} 
	\caption{Evolution of $\log(\max(|p_{\text{Rev-MEKF}} - p_{\text{truth}}|))$ depending on the logarithm of the noise in update step}
  \label{fig:evol_bias_position_imu_revmekf}
\end{figure}

The second observation concerns gyroscope bias. Rev-MEKF is able to compensate for such bias through its linear algebraic structure. As shown in Figure~\ref{fig:evol_bias_position_imu_revmekf}, the resulting precision can be made arbitrarily low depending on the chosen parameters of Rev-MEKF, whereas for MEKF, it will be bounded by the average acceleration variation during $dt$ equal to $10^{-1}m.s^{-2}$. Consequently, the final error is determined only by the accuracy of the accelerometer and magnetometer. 
However, one can observe that depending on the speed of computation of the bias, the first steps introduce an error in the position that is due to the linear approximation used in update step.

\begin{figure}[h!]
  \centering
  \includegraphics[width=0.33\textwidth]{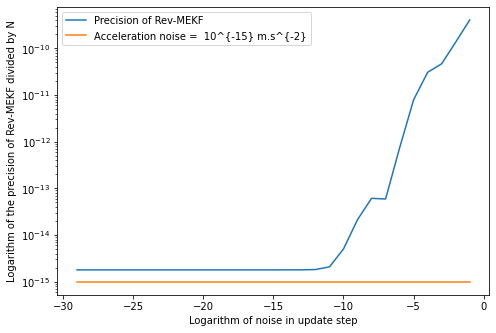} 
	\caption{Evolution of $\log(\max(|p_{\text{Rev-MEKF}} - p_{\text{truth}}|))$ depending on the logarithm of the noise in update step}
  \label{fig:evol_rnoise_revmekf}
\end{figure}

Third, in Figure~\ref{fig:evol_rnoise_revmekf}, one observes that the evolution of the precision depends only on the noise of the accelerometer, acting as an inferior limit on the precision established by Rev-MEKF, which is the expected behavior.

These experiments demonstrate that Rev-MEKF minimizes the loss of precision due to numerical approximation in MEKF implementations. A key conclusion is that, unlike MEKF, the error growth in Rev-MEKF depends solely on sensor noise and not on the type of motion. This makes the algorithm more robust and adaptable to extreme conditions.


\subsection{Real Data}

We conducted experiments under several scenarios using different sensor configurations, including odometry and IMU sensors. The main observation is that the Rev-MEKF performs well under idealized conditions. However, its usability in real-world scenarios is affected by several factors:
\begin{itemize}
    \item noise in the accelerometer and magnetometer measurements;
    \item errors in the computation of the gravity and magnetic field vectors.
\end{itemize}

As a result, the average distance between the finite set of rotations corresponding to the intersection of the acceleration vector with the surface (as defined in the previous section) is smaller than the noise induced by these factors. Consequently, distinguishing between two intersection points becomes impossible.

This average distance increases when the acceleration deviates from the gravity vector, i.e., during strong external accelerations, which correspond to situations where the MEKF performs poorly.

Therefore, Rev-MEKF must be adapted to apply linear-algebra-based corrections at appropriate instances. To this end, we introduce a detection step that determines when to switch between MEKF and Rev-MEKF, based on a trade-off between sensor noise and the type of motion. During static phases with noisy sensors, Rev-MEKF may perform worse than MEKF because the rotation angle $\alpha$, which projects an acceleration onto its closest point on the plane, can become large due to errors in the gravity vector and sensor noise. To mitigate this issue, we identify moments when the angle predicted by the Prediction step is closer to the angle returned by Rev-MEKF than to the angle corresponding to the raw accelerometer inclination. The procedure is detailed as follows:
\begin{itemize}
    \item To avoid errors due to global magnetic field computation, use
    \[
        q^{-1}*[0,0,1,0]*q
    \]
    instead of the raw magnetometer data, where $q$ is the predicted quaternion.
    \item By default, minimize the $z$ coordinate of the rotated acceleration, which is valid in most static phases.
    \item If the vector rotated in the linear algebra step never intersects the plane defined by the normal, or if the predicted rotation does not belong to the set of rotations that cross the plane, return the vector normally transmitted to MEKF.
    \item If the distance of the predicted rotation to the closest intersection rotation is $\gamma$ times smaller than the distance to the default rotation, return the vector normally returned by Rev-MEKF.
\end{itemize}

With this heuristic, the returned acceleration is significantly less noisy. We applied this approach to the dataset from~\cite{jeferson_menegazzo_2021}, produced with a vehicle driving on various road types over a long duration. The dataset was chosen because the sensors are well-calibrated and the acquisition duration allows meaningful comparisons between MEKF and Rev-MEKF.

\begin{figure}[h!]
  \centering
  \includegraphics[width=0.33\textwidth]{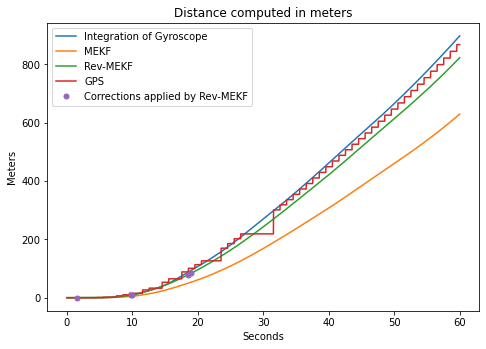} 
	\caption{Comparison of distances computed by different filter variants. Gyroscope noise: diagonal matrix of order $10^{-2}$; accelerometer and magnetometer noise: diagonal matrix of order $1$; $\gamma = 2$ and $\gamma=1.25$.}
  \label{fig:comparison_distance_imu}
\end{figure}

\begin{figure}[h!]
  \centering
  \includegraphics[width=0.33\textwidth]{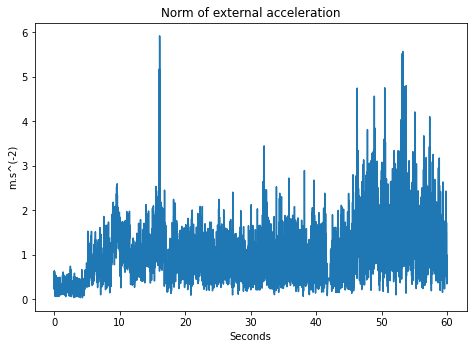} 
  \caption{Norm of the computed external acceleration}
  \label{fig:norm_acc_dataset}
\end{figure}

Figure~\ref{fig:comparison_distance_imu} illustrates the effect of the corrections on a straight trajectory on a planar surface. 
This case corresponds to a good scenario for our approach because the external acceleration is large, which is represented by the norm of the computed external acceleration in Figure~\ref{fig:norm_acc_dataset}, and the movement is on a surface almost horizontal.

\begin{figure}[h!]
  \centering
  \includegraphics[width=0.33\textwidth]{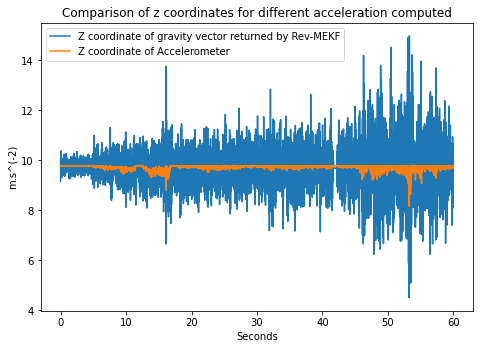} 
  \caption{Comparison of the $z$-coordinates of the gravity vector returned by the linear algebra step with the original accelerometer  $z$-coordinates.}
  \label{fig:comparison_acc_dataset}
\end{figure}

The most compelling effect is on the gravity vector transmitted to the update step, shown in Figure~\ref{fig:comparison_acc_dataset}, which demonstrates a noise reduction.

In this case, we have corrected less than $0.1\%$ of the whole global angular error produced by MEKF, with an average correction of $0.16$ radians.

It is important to carefully select $\gamma$ based on the application, plane normal, and typical accelerations. For instance, $\gamma = 1$ in the selected dataset introduces errors in static phases, counteracting the desired correction effect.

For odometry sensors, we also evaluated MEKF and Rev-MEKF. The effect on distance was less pronounced because the distance measurement relies on odometry rather than accelerometer data.
Moreover, the high external acceleration does not hold into this context. The consequence is that the percentage of corrections is very low, and the integration of the corresponding angle is thus negligible. However, the our corrections still acts on the acceleration because we are able to remove some noise on the raw accelerometer.
On one example, the amount of samples on which we apply a correction is equal to $0.3\%$ of $2000$ samples, with a correction with angles (in radians) equal to $0.27$ radians in average.
However, since the position is computed using odometry in a context where wheel slip is likely, this explains why the correction is negligible.
\section{Discussions}
\label{sec:discussions}

\subsection{Testability of MEKF}
A primary use of Rev-MEKF is to assess the numerical precision of MEKF under extreme conditions. With arbitrary precision, one can observe a linear growth of numerical error over time, allowing the evaluation of different MEKF implementations.

This approach enables comprehensive testing of potential errors, such as incorrect initialization, left/right multiplication mistakes, or improper transition matrices. It also facilitates adaptation of MEKF to new scenarios, including alternative residuals or sensor sets.

\subsection{Correction of Real Data}
In real-world conditions, Rev-MEKF cannot be applied in its pure form. Instead, one defines conditions under which linear algebra corrections are applied, typically when external acceleration is significant. This strategy can reduce vibrations in accelerometer readings, especially vertical vibrations during horizontal motion.

Future work should focus on identifying when corrections are necessary and estimating their magnitude, balancing the correction of large external accelerations with the effects of vibrations and noise that resemble external acceleration but should be considered static.

\subsection{Applications}
Using a normal vector in the Kalman filter enables a wide range of applications. Vehicle localization using odometry and IMU sensors is a primary example. The strategy is not limited to planar motion; it only requires an approximate plane containing the trajectory points, allowing the normal vector to evolve over time. This can include movements such as falls or robotic arm motions.

Other promising applications include satellite or space trajectory computation, where gyroscope and accelerometer sensors are typically low-noise and trajectories can be modeled on manifolds.

Wearable applications are also feasible, including human motion analysis during running, walking, or cycling, where motion can be approximated as planar.

Submarine localization is another example, relevant to nondestructive inspection of ship hulls. In this scenario, magnetometers may be unusable, but pressure sensors can provide a plane normal to project the trajectory.

\appendix[Application on Odometry input]
\label{sec:odometry}
Rev-MEKF can be adapted to set of sensors using Odometry in the following way.

The implementation of a Kalman filter for Odometry input associated to a differential-drive robot uses as an input left and right wheels input, and corrects the orientation this the accelerometer. The framework of this approach can be found in~\cite{Hanna2015}.

The Odometry input is given by two scalars representing the angle of rotation of the left and the right wheel. There is a relation between the distance traveled by the wheel, which is linearly related to the angle of rotation, and the distance and orientation of the differential-drive robot:
$$d = \frac{d_L+d_R}{2}\ \text{and}\ \Delta \theta = \frac{d_L-d_R}{2d_w} $$
where $d_w$ is the distance of a wheel to the center of the robot.
Thus, one can describe the prediction step in Algorithm~\ref{alg:AlgOdoPred}.

\begin{algorithm}

\caption{EKF Prediction Step for Odometry sensor}\label{alg:AlgOdoPred}
\begin{algorithmic}[1]

	\Require Distance traveled by left and right wheel $d_L$ and $d_R$, Covariance $P_{k-1}$, Process noise $Q_k$, Time step $dt$
\Ensure Updated $\mathcal{S} = \log(q)$, position $\mathbf{p}$, $P_k$
\State $[d,d \theta] \gets [\frac{d_L+d_R}{2},\frac{d_L-d_R}{2d_w}]$
\State $\theta \gets \mathcal{S}[3]$ \Comment{z-axis rotation angle}
\State $\mathcal{S} \gets \mathcal{S} + [0,0,d\theta]$
\State $\mathbf{p} \gets \mathbf{p} +[d\cdot \cos(\theta +d\theta/2),d\cdot \sin(\theta +d\theta/2),0]$
\State Initialize $J \gets I_{6}$,$K \gets 0_{2 \times 6}$
\State $J_{6,1:2} \gets -d\cdot [\sin(\theta),\cos(\theta)]$
\State $K_{1,1:2} \gets [\sin(\theta),\cos(\theta)])$
\State $K_{2,1:6} \gets [-\frac{d}{2}\sin(\theta),\frac{d}{2}\cos(\theta),0,0,0,1]$

\State $P_k \gets J^\top \cdot P_{k-1} \cdot J + K^\top Q_k K$
	\State \Return $\mathcal{S}, \mathbf{p}, P_k$
\end{algorithmic}
\end{algorithm}

Let us consider $(\mathcal{O},\mathbf{A},\mathbf{P},\Vec{n})$, giving Odometry, accelerometer, pressure input, and the normal.
The conditions that are used are different, since we use the fact that
$$(\mathbf{p}+\mathbf{s}\cdot dt -\Vec{g}\cdot (dt)^2 +R^{-1}\cdot \mathbf{A})\cdot \Vec{z} = \mathbf{P} \ \text{and}\ R^{-1} \Vec{n} = \Vec{z}.$$
For having a finite number of solutions, one has to ensure that $\Vec{n} \neq \Vec{z}$, which means that the plan cannot be orthogonal to the gravity vector. It is coherent with this set of sensors, because it corresponds to the case where accelerometer and pressure sensors are not able to bring information on the orientation.


%

%







\bibliographystyle{IEEEtran}
\bibliography{bibtex/bib/bare_jrnl}
\end{document}